\documentclass{iopart}
\usepackage{iopams}
\usepackage{graphicx}
\usepackage{color}
\newcommand{\be}{\begin{equation}}
\newcommand{\ee}{\end{equation}}
\newcommand{\bea}{\begin{eqnarray}}
\newcommand{\eea}{\end{eqnarray}}

\newcommand{\s}{\sigma}

\begin{document}

\title{Bethe ansatz solution of the topological Kondo model}

\author{A Altland$^1$, B B{\'e}ri$^2$, R Egger$^3$, and A M Tsvelik$^4$ }

\address{$^1$~Institut f\"ur Theoretische Physik,  
Universit\"at zu K\"oln, Z\"ulpicher Str. 77, D-50937 K\"oln, Germany\\
$^2$~School of Physics and Astronomy, University of Birmingham, Edgbaston, Birmingham B15 2TT, UK\\
$^3$~Institut f\"ur Theoretische Physik, Heinrich-Heine-Universit\"at, D-40225 D\"usseldorf, Germany\\
$^4$~Department of Condensed Matter Physics and Materials Science, Brookhaven National Laboratory,  Upton, NY 11973-5000, USA}

\submitto{JPA}

\begin{abstract}  
Conduction electrons coupled to a mesoscopic superconducting island hosting Majorana bound states have been shown to display a topological Kondo effect with robust non-Fermi liquid correlations. With $M$ bound states coupled to $M$ leads, this is an SO($M$) Kondo problem, with the asymptotic high and low energy theories known from bosonization and conformal field theory studies. Here we complement these approaches by analyzing the Bethe ansatz equations describing the exact solution of these models at all energy scales. We apply our findings to obtain nonperturbative results on the thermodynamics of $M\rightarrow M-2$ crossovers induced by tunnel couplings between adjacent Majorana bound states.  
\end{abstract}


%\maketitle

\section{Introduction} 

\begin{figure}
\begin{center}
\includegraphics[width=8cm]{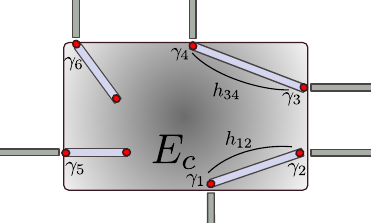}
\end{center}
\caption{\label{fig1} Schematic device setup for the topological Kondo effect with $M=6$ Majorana fermions $\gamma_j$ coupled to normal-conducting leads.  The Majorana bound states are realized as end states of spin-orbit coupled nanowires on a floating superconducting island with charging energy $E_c$.  The tunnel couplings $h_{j<k}$ between pairs of Majoranas act like a Zeeman field on the ``Majorana spin'' with components $i\gamma_j\gamma_k$.}
\end{figure}

Majorana fermion states are presently  among the most intensively 
studied objects in condensed matter physics. This is mainly due to the
non-Abelian anyon statistics of defects binding Majorana fermions, 
with promising applications to topological quantum information 
processing \cite{read2000paired,kitaev2001unpaired,FuKane08,sau2010generic,alicea2010majorana,oreg2010helical}.  
Majorana fermions are zero energy bound states, pairs of which 
form ``topological qubits" encoding ordinary fermion degrees of 
freedom in a non-local manner, 
see Refs.~\cite{mbsrev1,mbsrev2,mbsrev3} for recent reviews.
These exotic objects are predicted to arise in heterostructures 
combining simple $s$-wave superconductors and materials with 
strong spin-orbit coupling, and 
first experimental results on potential realizations based on  
semiconductor nanowires \cite{exp1,exp2,exp3,exp4,exp5} or 
topological insulators \cite{Williams2012,Knez2012} are under 
current investigation. 
Much of the work aimed at detecting
\cite{LawMaj,wimmer2011quantum,Futelep,Saucond,fidkowski,pientka2012enhanced} 
and manipulating
\cite{hassler2010anyonic,Saunet,alicea2011non,flensberg,vanheck2012coulomb,vanheckflux2013} Majorana fermions has so far been 
concerned with effectively noninteracting physics. 
However, as we have shown in a series of 
recent papers \cite{beri1,beri2,altland1,altland2,altland3,galpin}, Majorana fermions can also be a source of rich strongly correlated physics. 
Such a setup is realized by coupling  a ``floating'' mesoscopic 
superconducting island --- which has a finite charging energy $E_c$ ---
to normal-conducting lead electrodes via $M>2$ 
Majorana modes. The device setup is sketched in 
Fig.~\ref{fig1} and, in the parameter regime discussed below,
implies the ``topological Kondo effect'' \cite{beri1}.  
The Majorana fermions residing on the island thereby represent 
a quantum ``impurity'' which,
at low energy scales, becomes massively entangled with the 
conduction electrons in the attached leads through exchange-type processes.  
At low temperatures, this system is predicted to
  display exotic non-Fermi liquid correlations, similar to 
but different from the well-known overscreened multi-channel 
Kondo effect \cite{wiegmann,andrei,tsvelik,AL}.
While achieving such non-Fermi liquid correlations in conventional
Kondo devices is usually hindered 
by the competition of various couplings and, in particular, 
by the fact that channel anisotropy
is a relevant perturbation destroying the non-Fermi liquid 
fixed point \cite{gogolinbook}, the couplings acting 
against the topological Kondo effect can be eliminated to exponential 
accuracy simply by ensuring that adjacent Majorana fermions are 
sufficiently far apart. 
We note in passing that related physics has also been 
predicted in junctions of transverse Ising spin 
chains \cite{tsvelikprl,crampe,tsveliknjp}. However, despite of
superficial similarities, the topological Kondo effect found in
the Majorana device in Fig.~\ref{fig1} is substantially 
different and always characterized by non-Fermi liquid behavior.

The source of the topological Kondo effect is the combination of the island's 
charging energy, $E_c$, and the presence of the topological qubit degrees
 of freedom associated to the Majoranas. The former ensures that the
 island has a definite number of electrons in its ground state, while 
the latter transforms the island into an effective non-local ``Majorana spin'' 
at energies much below both $E_c$ and the gap to non-Majorana 
(quasi-particle) excitations. 
In this work we only discuss the physics on those low energy scales.
When conduction electrons in a given lead electrode (labeled
by the index $j$) have a finite tunnel coupling, $t_j$, 
to this non-local Majorana spin via the Majorana fermion $\gamma_j$, 
the screening correlations ultimately responsible for non-Fermi liquid behavior arise.

As the $M$ Majorana fermions  are described by operators
$\gamma_j=\gamma_j^\dagger$ subject to the Clifford algebra $
\{\gamma_j,\gamma_k\} = 2\delta_{jk}$ \cite{mbsrev1,mbsrev2,mbsrev3},
the relevant symmetry group for this Kondo problem is SO($M$). 
Indeed, the $\{ \gamma_j \}$ compose a spinor representation of 
the SO($M$) group, where the $M(M-1)/2$ different products 
$i\gamma_j\gamma_k$ represent the different components of the
``bare'' (i.e., uncoupled to leads) Majorana spin \cite{altland3}. 
The Kondo limit is realized when the charging energy $E_c \gg {\rm max} 
(t_j^2/v)$; in what follows we set the  Fermi velocity
$v=1$ and use units with $\hbar=k_B=1$. 
The effective Hamiltonian describing the system at low energy 
scales, i.e., below $E_c$ and the energy of non-Majorana excitations,
is  \cite{beri1,beri2,altland1,altland2}
\begin{eqnarray} \label{model}
H &=& -i\sum_{j=1}^M \int_{-\infty}^\infty dx \ \psi_j^\dagger(x) 
\partial_x \psi_j^{}(x) \\ &+&\nonumber \sum_{j\neq k} \lambda_{jk}
\gamma_{j} \gamma_{k} \psi_{k}^{\dagger}(0) \psi^{}_{j}(0)
+ i\sum_{j\neq k} h_{jk}\gamma_j\gamma_k.
\end{eqnarray}
Here $\psi_{j}(x)$ is an effectively spinless right-moving fermion field
describing the $j$th lead, where we unfolded from the physical lead 
coordinates $x<0$ to the full line;  $x=0$ is the coordinate of the
 tunnel contact.  
The symmetric matrix $\lambda_{jk}\approx t_j t_k/E_c >0$ is the 
analogue of the exchange coupling in the usual Kondo problem.  
The non-local couplings $h_{jk}=-h_{kj}$ correspond to overlaps 
between Majorana bound states. 
While they can in principle be suppressed to exponential accuracy 
by separating the Majoranas from each other, we include them here for 
two reasons. First, instead of suppressing these couplings, 
they can be deliberately switched on (by applying gate voltages)
and used as handles to probe
the physics of the Kondo screened non-local SO($M$) spin \cite{altland3}.
Second, their inclusion also allows one to study the eventual 
fate of the non-Fermi liquid physics at the lowest energy scales 
\cite{altland3,galpin}.
Note that in the Kondo language, the couplings $h_{jk}$ act like Zeeman fields. 
We mention in passing that the $M$ Majoranas appearing in Eq.~(\ref{model}) 
might only be a subset of all Majorana bound states 
on the island. Regarding the complementary Majoranas which are not 
coupled to any lead electrode, we assume that these have no direct tunnel 
couplings with the $\gamma_j$.  

In the absence of the Zeeman field,
the weak-coupling renormalization group (RG) analysis for Eq.~(\ref{model}) 
indicates a flow of the exchange couplings towards
a strong-coupling isotropic limit \cite{beri1,beri2,altland1,altland2}.
Taking all exchange couplings equal, $\lambda_{j\ne k} = \bar\lambda$, 
the perturbative RG approach turns out to break down on energy scales
below the Kondo temperature  
\begin{equation}\label{tk}
T_K\simeq E_c \exp\left(- \frac{\pi}{(M-2) \bar \lambda}\right).
\end{equation}
For temperatures $T\ll T_K$, one enters the non-Fermi liquid regime
of the topological Kondo effect corresponding to an SO$_2(M)$ 
Wess-Zumino-Novikov-Witten boundary conformal field theory (BCFT)
\cite{AL,CFT}.
Crucially, anisotropy in the $\lambda_{jk}$ is an irrelevant perturbation
around this fixed point, in contrast to conventional overscreened multi-channel Kondo fixed points. 
However, anisotropy is likely to break integrability for $M>4$.

The asymptotic low and high energy physics of the topological 
Kondo effect has been studied in Refs.~\cite{beri1,beri2,altland1,altland2} for 
$h_{jk}=0$, while the effects of the Zeeman field components $h_{jk}$ 
have been addressed in Ref.~\cite{altland3}. 
For the $M=3$ case, Ref.~\cite{galpin} has explored the full crossover from 
high to low energies,  with and without Zeeman fields, using the 
numerical renormalization group technique. Here we complement those 
previous studies by providing  analytical results obtained from exact
Bethe ansatz (BA) calculations.  Note that BA results for related spin chain junctions can be found in Ref.~\cite{tsveliknjp}.
Before turning to the detailed BA solution of the topological Kondo
model (\ref{model}), we first summarize the main results of this paper.

\section{Summary of results}\label{sec2}

In principle, the BA solution allows one to 
obtain very detailed information about the system, 
including the entire energy spectrum (for finite-length leads). This can provide access to the full thermodynamic information 
for arbitrary temperature and Zeeman field. 
The BA solution certainly exists for $M=3$, where Eq.~(\ref{model}) is 
equivalent to the four-channel Kondo model with 
spin $S=1/2$ \cite{Fabrizio1994}. 
For general $M$, we make a plausible conjecture for  the BA and justify it 
by running various checks.  In particular, we verify that the BA 
reproduces the results obtained by means of BCFT, 
including the ground-state impurity entropy. 
As a concrete application, we will discuss the 
thermodynamic response of the topological Kondo system 
to a Zeeman field by computing the ``magnetization'', i.e., the expectation value of the Majorana spin, 
\begin{equation} \label{magnetization}
{\cal M}_{jk}=\langle i\gamma_j\gamma_k\rangle,
\end{equation} 
for which a measurement scheme  has been proposed in Ref.~\cite{altland3}.
In particular, we show that in the presence of just one 
Zeeman field component, say $h_{12}> 0$, a crossover 
between a Kondo problem with SO($M)$ symmetry to another one
with the symmetry group SO($M-2$) is induced upon lowering the
temperature.  For $M\le 4$, instead of another Kondo fixed point,
the $M\to M-2$ crossover terminates at a Fermi liquid state \cite{egger2}. 
The BA solution discussed below allows one to address this
crossover in a nonperturbative fashion. 
This is an attractive feature since the Zeeman field 
is an RG-relevant perturbation destabilizing 
the topological Kondo fixed point on temperatures below
the temperature scale
\be
T_h \simeq T_K ({h_{12}}/T_K)^{M/2} , \label{Th}
\ee
which follows from simple dimensional scaling arguments \cite{altland3}.
Effectively, for $T\ll T_h$, the Majorana modes $\gamma_1$ and $\gamma_2$
coupled by the Zeeman field move to finite energy and thus disappear
from the low-energy theory.  The remaining $M-2$ 
Majoranas then realize a non-Fermi liquid SO$_2(M-2)$ Kondo fixed point
(assuming $M>4$).
In this flow, the Kondo temperature  $T_K$ 
represents the high-energy cutoff, while the ``effective" Kondo temperature 
for the emergent low-temperature fixed point of SO$_2(M-2$) symmetry is
set by $T_K^{(M-2)}=T_h$.  
Our BA solution, explicitly constructed for $M=3$ up to $M=6$ below, nicely confirms this intuitive picture.

For each of the considered $M$, we also compute the ground-state impurity entropy $S_{\rm imp}$.  In the absence of the Zeeman field, we find that a BCFT calculation yields 
\begin{equation}\label{gsent}
S_{\rm imp}=\ln d_M,\quad d_M=\left\{ \begin{array}{ll} \sqrt{M},&  M\ \textrm{odd},\\ \sqrt{M/2}, & M\ \textrm{even}\end{array} \right.
\end{equation}
The quantum dimension $d_M$ thus strongly responds to the parity of $M$. This BCFT result follows from arguments very similar to those in Ref.~\cite{oshikawa}, and the perfect agreement with our BA results provides a consistency check for the latter. 

Let us start with the case $M=3$, where the model 
(\ref{model}) maps \cite{Fabrizio1994} to the four-channel SU(2)
 Kondo problem with impurity spin $S=1/2$. This correspondence allows us 
to use BA results obtained for the latter model \cite{wiegmann,tsvelik}. 
In particular, the ground-state magnetization ${\cal M}_{12}$
can be expressed as scaling function of the variable
$Y=h_{12}/\kappa T_K$, where $\kappa=\frac{64\pi}{e^2}\simeq 27.21$ and $T_K$
is given by Eq.~(\ref{tk}).  For $Y\ll 1$ and $T=0$, we find
\begin{equation} \label{magn3}
{\cal M}_{12} \simeq 0.868 \sqrt{Y} + 0.034 Y \ln Y  + O\left( Y^3 \right).
\end{equation}
 The case $M=3$ is the one where the singularities in the 
thermodynamic quantities and correlation functions are the strongest.
Indeed, the susceptibility, $\chi_{12}=\partial {\cal M}_{12}/\partial h_{12}$, has a $Y^{-1/2}$ singularity for $Y\to 0$.  The $M\to M-2$ crossover picture here turns out to be consistent with a $T=0$ fixed point describing a conventional Coulomb blockade regime, where the island ultimately decouples from the environment. 

For $M > 4$, the $M\to M-2$ crossover induced by lowering the temperature instead terminates in another Kondo fixed point with symmetry group SO$(M-2$).  In this case, the breakdown of perturbation theory at
low temperatures $T\ll T_h$ signals in a weak singularity. For instance, for $M=6$, we obtain  a
logarithmically divergent second derivative of ${\cal M}_{12}$,
\be
\frac{\partial^2{\cal
M}_{12}}{\partial h^2_{12}} \sim T_K^{-2}\ln(T_K/h_{12}). \label{nonlinear}
\ee
To observe the flow $M \to M-2$  we 
turn on  an additional weak field component $h_{34}\ll h_{12}$. Then at $h_{34} \ll T_{h}$, already the
first derivative of the corresponding magnetization component diverges,
\be \label{eq:SU22}
\frac{\partial {\cal M}_{34}} {\partial h_{34}} \sim
T_h^{-1}\ln(T_h/h_{34}), 
\ee
in a manner well known for the two-channel SU$(2)$
Kondo model \cite{wiegmann,tsvelik}. This behavior can be understood from the equivalence of the SO(4) and two-channel SU$(2)$ Kondo fixed points \cite{beri2}, which is also recovered from the structure of the BA equations. 

The rest of the paper is organized as follows. We begin by discussing the problem for $M=3$ and $M=6$, 
including a study of the $M\to M-2$ crossover in the latter case. 
In these cases, as well as for $M=4$, the BA analysis is aided by the equivalence  relations
\begin{equation}\label{equivalence}
{\rm SO}(3)\sim {\rm SU}(2), \quad {\rm SO}(6)\sim {\rm SU}(4), \quad
{\rm SO}(4)\sim {\rm SU}(2)\times {\rm SU}(2), 
\end{equation}
which establish links to previously studied SU($N$) Kondo models \cite{wiegmann,andrei,tsvelik,jerez}. We will then combine our $M=6$ equations with general results, linking the group algebra to the structure of the BA equations, to suggest a generalization for arbitrary even $M$. We also propose the corresponding equations for odd $M>3$, and discuss the $M=5\to 3$ crossover. This case (unlike $M=3,4,6$) does not correspond to impurity models previously studied in the SU($N$) Kondo context. 

\section{Bethe ansatz solution}

The strategy pursued below is as follows.
The general classification scheme of possible BA equations for models with Lie group symmetry is known. This allows us to address the problem with equal coupling constants $\lambda_{jk} = \bar\lambda$; we thus assume  isotropic exchange couplings unless stated otherwise.
We consider finite length $L$ of the leads with periodic boundary conditions, and take the thermodynamic limit afterwards. According to  Refs.~\cite{reshetikhin} and \cite{PCF}, the general form of the BA equations is then dictated by the Dynkin diagram of the corresponding group.  The next step is to determine the position of the so-called driving terms and their precise form. These are determined by representations of the bulk Hamiltonian and the impurity spin.
We also check our BA equations against previously known results. 

\subsection{Case $M=3$}

For $M=3$, the model (\ref{model}) is equivalent to the four-channel SU(2) Kondo problem with impurity spin $S=1/2$ \cite{Fabrizio1994}. To see this equivalence, let us introduce the operators 
$J_j = \frac{i}{4}\varepsilon_{jkl}\gamma_k\gamma_l$ (with $j=1,2,3=x,y,z$ and summation convention), 
which are equivalent to the components of a
spin $S=1/2$ operator \cite{martin}. 
Then the exchange interaction in Eq.~(\ref{model}) reads 
\begin{eqnarray}
 H_K &=& 4i\lambda_{12} J_x\psi^\dagger_1(0)\psi_2(0)  + 
4i \lambda_{23} J_y\psi^\dagger_2(0)\psi_3(0) \nonumber \\
 &+& 4i \lambda_{13} J_z \psi^\dagger_3(0)\psi_1(0)  + \textrm{h.c.}
\end{eqnarray}
Expanding the bulk fermions  in their real and imaginary Majorana components,
 $\psi_j(x) = \chi_j(x) + i\xi_j(x)$, we get 
$i\varepsilon_{jkl}(\psi^\dagger_k\psi_l - \psi^\dagger_l\psi_k) =
i\varepsilon_{jkl}(\chi_k\chi_l + \xi_l\xi_k)$, i.e.,  
the sum of two 
%SU$_2$(2) 
SO$_1(3)$ currents equals the SU$_4$(2) current. 
The result is the anisotropic spin $S=1/2$ four-channel 
Kondo model, which is integrable.
The BA solution for this model has been thoroughly studied 
\cite{wiegmann,tsvelik}, and thereby applies also to the SO(3) topological
Kondo effect. In particular one finds Eq.~(\ref{magn3}) for the magnetization, and $S_{\rm imp}=\ln \sqrt{3}$  
in the absence of the Zeeman field.

\subsection{Case $M=6$}

Next we discuss the case $M=6$. We demonstrate that a
Zeeman field coupling just one pair of Majoranas drives the system from $M=6$ to $M=4$, which in turn is equivalent to the SU$_2$(2) Kondo effect \cite{beri2}. 
We restrict our analysis to Zeeman fields $h_{jk}$ that couple to commuting  pairs of Majoranas forming a Cartan subalgebra. These pairs could be chosen arbitrarily but should not overlap.  For $M$ wires, we then have $[M/2]$ pairs ($[x]$ is the integer part of $x$). For $M=6$, the non-vanishing Zeeman couplings are taken as 
\be \label{Zeeman2}
H_1=h_{12},\quad H_2=h_{34},  \quad H_3=h_{56}.´
\ee

For $M=6$, the BA equations have the same general form as for the 
SU$_2$(4) Kondo model because the corresponding Dynkin diagrams coincide. Since creation and annihilation  operators of the  bulk fermions transform according to the vector representation of the O(6) group, this representation is isomorphic  to the representation of the SU(4) group,
where the Young tableaux have one column containing two boxes.  The suggested  BA equations are, with rapidities $x_a^{(j)}$, given by   
\bea
&& e_1(x_a^{(1)} -1/\bar\lambda)
\prod_{b=1}^{M_2}e_{1}(x_a^{(1)} -x_b^{(2)}) = 
\prod_{b=1}^{M_1}e_2(x_a^{(1)} -x_b^{(1)}),\\ \nonumber
&& \left [e_2(x_a^{(2)})\right]^N \prod_{b=1}^{M_2}e_{1}
(x_a^{(2)} -x_b^{(1)})
\prod_{b=1}^{M_3} e_{1} (x_a^{(2)} -x_b^{(3)} ) =
 \prod_{b=1}^{M_2}e_2(x_a^{(2)} -x_b^{(2)}), \\ \nonumber
&& \prod_{b=1}^{M_2}e_{1}(x_a^{(3)} -x_b^{(2)}) = 
\prod_{b=1}^{M_3}e_2 (x_a^{(3)} -x_b^{(3)}), 
\eea
where the energy follows as
\begin{equation}
E = \sum_{a=1}^{M_2} \frac{1}{2i}\ln e_2(x_a^{(2)}),\quad
e_n(x) = \frac{x- in/2}{x+in/2}.
\label{bareBA}
\end{equation}
Here $N$ is the number of particles in the Fermi sea, and $M_{1,2,3}$ are integer numbers equal to linear combinations of eigenvalues of the Cartan operators of the group \cite{PCF}.  
The driving term for the bulk is located in the second equation, as it 
fits the vector representation of the O(6) group.  
Since the impurity spin is in the spinor representation, its driving term is in the first equation.  

As a first step in the derivation of the thermodynamic Bethe ansatz (TBA) equations, we then classify solutions of the bare 
BA equations (\ref{bareBA}).  These solutions are complex, but in the 
thermodynamic limit, $L \to \infty$ with $N/L$ and  $M_{1,2,3}/L$ finite,  their imaginary parts are simple:  They group into clusters with 
common real part $X_{\alpha}^{(j,n)}$, the so-called 'strings', where the
rapidities are given by 
\begin{equation}
 x_{n,p;\alpha}^{(j)} = X_{\alpha}^{(j,n)} + 
\frac{i}{2}(n+1-2p) + O\left(e^{-c_0L}\right), 
\end{equation}
with $n=1,2,\ldots$,  $p = 1,\ldots,n$, and $c_0>0$. 
As next step, we introduce distribution functions for
rapidities of string centers, $\rho_n^{(j)}(x)$, 
and unoccupied spaces, $\tilde\rho_n^{(j)}(x)$. 
The discrete equations (\ref{bareBA}) are thereby transformed into 
integral equations relating $\tilde\rho$ and $\rho$,
\bea
&& \tilde\rho_n^{(j)} + A_{nm}*C^{jl}*\rho_m^{(l)} =  A_{n,2}*s(x) \delta^{j,2} + 
\frac{1}{N}a_n(x-1/\bar\lambda)\delta^{j,1}, \label{dens}
\eea
where $n = 1,2,\ldots$, $j=1,2,3$, we use the summation convention, convolutions are denoted by a star, i.e.,
$f*g(x) = \int dy f(x-y)g(y)$, and 
\bea
A_{nm}(\omega) &=& \coth(|\omega|/2) \left( e^{-|n-m||\omega|/2} - 
e^{-(n+m)|\omega|/2}\right), \nonumber\\
C_{nm}(\omega) &=& \delta_{nm} - s(\omega) \left(\delta_{n,m-1} + 
\delta_{n,m+1}\right), \nonumber\\
s(\omega) &=& \frac{1}{2\cosh(\omega/2)}, \quad
 a_n(\omega) = e^{-n|\omega|/2}. \label{C}
\eea
The TBA equations now follow by minimization of the generalized 
free energy, $F=E-TS$, subject to the constraints imposed by Eq.~(\ref{dens})
and with the entropy  
\bea \nonumber
S &=& N\sum_{n=0}^{\infty}\sum_j\int d x 
\Big[ (\rho_n^{(j)} + \tilde\rho_n^{(j)})\ln(\rho_n^{(j)} + 
\tilde\rho_n^{(j)})  \\
&& - \rho_n^{(j)}\ln\rho_n^{(j)} - \tilde\rho_n^{(j)}
\ln\tilde\rho_n^{(j)}\Big].
\eea
The TBA equations determine ratios of the distribution functions, 
which are collected in $\phi_n^{(j)}$ functions
according to 
\be 
\tilde\rho_n^{(j)}(x)/\rho_n^{(j)}(x) = e^{-\phi_n^{(j)}(x)}.
\ee
For $M=6$, the TBA equations in the scaling limit coincide with  
those of the SU$_2(4)$ Kondo (or Coqblin-Schrieffer) model,
 with the impurity in the fundamental (single box) representation \cite{jerez},
\bea
&& F_{\rm imp} = - T\sum_{j=1}^3\int dx f_j \left[x + \frac{\ln(T_K/T)}{\pi}\right]
\ln\left(1 +e^{\phi_1^{(j)}(x)}\right), \nonumber\\
&& \ln\left(1 + e^{-\phi_n^{(j)}}\right) - {\cal A}_{jl}*C_{nm}*\ln\left(1 + 
e^{\phi_m^{(l)}}\right)= \\ \nonumber
&&=  \delta_{n,2}\sin(\pi j/4) e^{-\pi x/2} ,
\eea
where $j,l=1,2,3$ and $n,m=1,2,\ldots$. 
The Zeeman fields $H_j$ in Eq.~(\ref{Zeeman2}) enter through the 
constraint
\begin{equation}
\lim_{n\to \infty} \frac{\phi_n^{(j)}}{n} = H_j/T,
\label{F6}
\end{equation}
 and the Fourier transforms of the above kernels are given by
\bea
&&  f_j(\omega) = \frac{\sinh[(2-j/2)\omega]}{\sinh(2\omega)}, \quad {\cal A}_{jl} = [C^{-1}]_{jl} = \nonumber\\
&&  =2\coth(\omega/2)\frac{\sinh[(2 - \mbox{max}(j,l)/2)\omega]
\sinh[\mbox{min}(j,l)\omega/2]}{\sinh (2\omega)}.
\eea

The ground-state  impurity entropy is then determined by the 
asymptotics of $\phi_1^{(j)}(-\infty)$.
In the absence of the Zeeman field,
the solution for the general SU$_k(N)$ case is 
\bea
1+e^{\phi_n^{(j)}(-\infty)} = \frac{\sin\left[\frac{\pi(n+ N-j)}
{k+N}\right]\sin\left[\frac{\pi(n+j)}{k+N}\right]}{\sin\left
[\frac{\pi(N-j)}{k+N}\right]\sin\left[\frac{\pi j}{k+N}\right]}.
\eea
Substituting this into the above equations
and putting $n=1,$ $N=4,$ and  $k=2$, as is appropriate for our SO$_2(6$) problem [see Eq.~(\ref{equivalence})], we obtain $S_{\rm imp} = \ln\sqrt 3$ in accordance with Eq.~(\ref{gsent}). 

Below we consider the thermodynamics at $T=0$, such that 
the equations for the ground-state root densities suffice. 
All roots of the BA equations are in $n=2$ strings of 
different 'colors', $j =1,2,3$. 
In that case, Eqs.~(\ref{dens}) are reduced to a set of Wiener-Hopf equations, 
\bea
&& s(x)\delta^{j,2} + 
\frac{\delta^{j,1}}{N}[s*s](x-1/\bar\lambda) 
=  [A_{2,2}]^{-1}\tilde\rho^{(2)}_{2}(x) + 
C_{jl}*\rho^{(l)}_{2}(x). \label{dens2}
\eea
Let us then isolate the terms proportional to $1/N$, which are associated with the impurity,
\begin{equation}
\rho = \rho_{b} + \frac{1}{N}\rho_{\rm imp},
\end{equation} 
where $\rho=\rho_2$ or $\tilde\rho_2$. 
In the ground state, the $j$th-order roots fill the 
interval $(-\infty, B_j)$, where the limits $B_j$ are determined by 
the Zeeman fields $H_j$ in Eq.~(\ref{Zeeman2}),
\begin{equation}
\chi H_j = \int_{B_j}^{\infty} d x [\tilde\rho^{(j)}(x)]_b,
\end{equation}
with the bulk susceptibility $\chi=1/(2\pi)$  (we use $v=1$). 
To proceed from this point on, we need to specify the precise
 Zeeman field configuration. 

\subsubsection{All $B_j$ equal.}

With Eq.~(\ref{Zeeman2}), we consider the Zeeman field configuration with 
\begin{equation}
H_j = h_0\sin(\pi j/4),\quad j=1,2,3,
\end{equation} 
where all $B_j$ are equal.
In this case, Eqs.~(\ref{dens2}) describe the vicinity of 
the SU$_2$(4) fixed point for small $h_0$.
The system of Wiener-Hopf equations (\ref{dens2}) with equal limits of integration is then solved by
\bea 
&& \rho^{(j),+}(\omega=0) =\frac{1}{16\pi i}\sum_{l=1}^3\sin(\pi jl/4) \\ \nonumber &&\times \int\frac{d\omega}{\omega -i 0^+}\frac{f_l^{(-)}(\omega) \exp\left[\frac{2i\omega}{\pi}\ln\left(h_0/ \left[f_1^{(-)}(-i\pi/2) T_K\right]\right)\right]}
{2\cosh(\omega/2)[\cosh(\omega/2)-\cos(\pi l/4)]},\\
&& f_l^{(-)}(\omega) =\frac{ \left(  \nonumber
\frac{i\omega +0^+}{\pi e}\right)^{i\omega/\pi}
\left[\left(\frac{\omega -i0^+}{4\pi}\right)^2 + 
(l/8)^2\right]^{1/2}}{\Gamma(\frac12 +i \frac{\omega}{2\pi})
\Gamma(1 - \frac{l}{8 }+ i\frac{\omega}{4\pi})
\Gamma(1 +\frac{l}{8} + i\frac{\omega}{4\pi})},
\eea
with the Gamma function $\Gamma$. 
 For small Zeeman fields, $h_0 \ll T_K$,
 the result is dominated by the linear term but acquires 
a non-Fermi liquid correction,
\bea
\rho^{(j),+}(\omega =0) = \sin(\pi j/4)\frac{h_0}{2\pi T_K} + b_j \frac{h_0^2}{T_K^2} \ln(T_K/h_0) + \cdots,
\eea
where $b_j$ is a numerical coefficient.  The second (non-Fermi liquid) term originates from the double pole in the integrand at $\omega = -i\pi$.
From here on it is straightforward to 
obtain Eq.~(\ref{nonlinear}) for the magnetization, which has
been quoted in Sec.~\ref{sec2}.

\subsubsection{Case $H_1\gg H_{2,3}$ and $M\to M-2$ flow.}

Consider next a Zeeman field where the amount of holes (unoccupied spaces) 
in the $j=2$ equation, see Eq.~(\ref{dens2}), strongly exceeds their amount in the $j=1,3$ equations,
such  that $B_2 \ll B_{1,3}$. This situation is realized when one of the Zeeman field components by far exceeds the others, for instance, $H_{1} \gg H_{2}, H_{3}$ in Eq.~(\ref{Zeeman2}). 
The $H_1$ field then generates the temperature scale $T_h$ in Eq.~(\ref{Th}), below which the physics 
is expected to be determined by the SO(4)~$\sim$~SU$_2$(2) Kondo effect.
In this limit, we can neglect $\tilde\rho^{(1,3)}$
and rewrite Eq.~(\ref{dens2}) as  
\bea
&& K*\tilde\rho^{(1)} + \rho^{(1)} = s*\rho^{(2)} +  \frac{1}{N}[s*s](x
-1/\bar\lambda),\label{M2}\\
&& K*\tilde\rho^{(3)} + \rho^{(3)} = s*\rho^{(2)}, \label{M2a}\\
&& \rho^{(2)} +K*{\cal A}_{2,2}*\tilde\rho^{(2)} = 
{\cal A}_{2,2}*s +  \frac{1}{N}{\cal A}_{2,2}*[s*s](x-1/\bar
\lambda), \label{dens3}\\ \nonumber
&& K(\omega) = [A_{2,2}]^{-1}=\frac{1}{1+ e^{-2|\omega|}}.
\eea
The densities $\tilde\rho^{(3)}$ and $\rho^{(3)}$ do not contribute to the impurity thermodynamics. In the scaling limit, we have to keep only the asymptotics of the 
bulk driving term, and the explicit form of Eq.~(\ref{dens3}) is 
\bea
&& \rho^{(2)}(x) + \int_{B_2}^{\infty} d y K(x-y)\tilde\rho^{(2)}(y) =\nonumber \\ && =
\frac{1}{\sqrt 2}e^{-\pi x/2} +  \frac{1}{N} \frac{1}{2\cosh[\pi(x- 1/\bar\lambda)/2]}. \label{WH}
\eea  
Equations (\ref{M2}) and (\ref{M2a}) determine the magnetization components ${\cal M}_{jk}$ with $(j,k)\ne (1,2)$, which are not directly affected by the large Zeeman field $H_1=h_{12}$. 
Since $B_2 \ll B_{1,3}$, one can approximate $s*\rho^{(2)}$ by  
the asymptotic expression 
\bea \nonumber
&& s*\rho^{(2)} \approx (A + \frac{1}{N}A') e^{-\pi x}, \\  
&& A + \frac{1}{N}A'= \int_{-\infty}^{B_2} d y e^{\pi y}\rho^{(2)}(y). 
\label{asymp}
\eea
Then Eq.~(\ref{M2}) coincides with the equation for the ground-state 
root density of the SU$_2$(2) Kondo model, and Eq.~(\ref{M2a}) coincides  
with the equation for the SU$_1$(2) Kondo model in the Fermi liquid limit. 
Indeed, the impurity part $\sim 1/N$ of Eq.~(\ref{asymp}) is also 
$\sim e^{- \pi x}$, and we have 
\bea
&& K*\tilde\rho^{(1)} + \rho^{(1)} =   A 
e^{-\pi x} + \frac{1}{N}[s*s](x-1/\bar\lambda),\label{M2b}\\
&& K*\tilde\rho^{(3)} + \rho^{(3)} = A e^{-\pi x} + 
\frac{1}{N}A' e^{-\pi x}. \label{M2ab}
\eea
Eliminating the prefactor $A$ in Eq.~(\ref{M2b}) by a shift of $x$,
we bring Eq.~(\ref{M2b}) to the canonical form for the SU$_2$(2) model,
with the \textit{renormalized Kondo temperature} for the emergent $M=4$ Kondo model, 
\be
T_K^{(4)} = A^{-1}e^{-\pi/\bar\lambda}. \label{TK2}
\ee
The factor $A$ can now be extracted from the solution of the
Wiener-Hopf equation (\ref{WH}),
\bea
&& \rho^{(+)}_2(\omega) = \frac{2}{\pi G^{(-)}
(-i\pi/2)G^{(+)}(\omega)}e^{-\pi B_2/2} + \\
&& \frac{1}{N}\frac{i}{2\pi G^{(+)}(\omega)}\int 
\frac{d \omega'}{\omega'-\omega +i 0^+}\frac{e^{-i\omega'(B_2-1/\bar\lambda)}}
{2\cosh(\omega') G^{(-)}(\omega')}, \nonumber
\\ \nonumber && G^{(+)}(-\omega) = G^{(-)}(\omega) = 
\frac{\Gamma(1/2 + i\omega/\pi)}{\sqrt\pi}\left(
\frac{i\omega- 0^+}{\pi e}\right)^{-i\omega/\pi}.
\eea 
The bulk part of this expression yields the bulk 
magnetic moment $\chi H_1$, and thus determines the value of $B_2$.
In fact, we find
\be 
\chi H_1 =\frac{ \sqrt 8 e^{-\pi B_2/2}}{\pi G^{(-)}(-i\pi/2)}.
\ee 
Substituting this into Eq.~(\ref{asymp}),  we get $A = (\chi H_1)^3/\pi$.  
We now plug this result back into Eq.~(\ref{TK2}) and take into account that the $M=6$ Kondo temperature $T_K^{(6)}$ (defined for $h_{jk}=0$) is  
given by Eq.~(\ref{tk}).  With an extra factor two in the 
exponent because of the different normalization of SO(6) and SU(4) 
generators, the $M=6$ Kondo temperature here reads
\be
T_K^{(6)} = \chi^{-1}\exp\left(-\frac{\pi}{2\bar\lambda}\right),
\ee
and therefore we finally get the Kondo temperature of the 
effective low-energy SO($4$) model, realized at $T\ll T_h$, in the form 
\bea
T_K^{(4)} = \frac{1}{\pi} T_K^{(6)}\left(H_1/T_K^{(6)}\right)^3.
\eea
Up to a prefactor of order unity, this scale coincides with
the crossover scale $T_h$ in Eq.~(\ref{Th}).
We have thus shown that $T_h$ acts as the Kondo temperature
for the emergent SO($M-2)$ topological Kondo effect.

The well-known result for the magnetization of the two-channel Kondo model \cite{wiegmann}, 
\[
\langle S^z\rangle  \sim \frac{H_1}{2\pi T_K}\ln(T_K/H_1),
\]
then dictates the $T=0$ magnetization behavior announced
above in Eq.~(\ref{eq:SU22}).  Moreover, 
the results of Ref.~\cite{tsvelik} yield the ground-state 
impurity entropy for the SU$_2$(2) case corresponding to $M=4$, 
$S_{\rm imp} = \ln\sqrt 2$, which is again in agreement with Eq.~(\ref{gsent}).

Now that we have successfully run the checks for $M=6$, we can 
write down the BA equations for arbitrary even $M$. 
Putting $M=2K$, they read
 \bea
&& e_1(x_a^{(K-1)} -1/\bar\lambda)
\prod_{b=1}^{M_{K-2}}e_{1}(x_a^{(K-1)} -x_b^{(K-2)}) \nonumber =\\ && 
\prod_{b=1}^{M_{K-1}}e_2(x_a^{(K-1)} -x_b^{(K-1)}),\nonumber\\
&& \prod_{b=1}^{M_{K-1}}e_{1}
(x_a^{(K-2)} -x_b^{(K-1)})
\prod_{b=1}^{M_K}e_{1}(x_a^{(K-2)} -x_b^{(K)})\\
\nonumber && \prod_{b=1}^{M_{K-3}}e_1(x_a^{(K-2)} -x_b^{(K-3)}) = \prod_{b=1}^{M_{K-2}}e_2(x_a^{(K-2)} -x_b^{(K-2)}), \\ \nonumber
&& \prod_{b=1}^{M_{p-1}}e_1(x_a^{(p)} - 
x_b^{(p-1)})\prod_{b=1}^{M_{p+1}}e_1(x_a^{(p)} - 
x_b^{(p+1)})\\ \nonumber &&
 = \prod_{b=1}^{M_p}e_2(x_a^{(p)} - x_b^{(p)}), \quad p=2,\ldots,K-1,\\
&& [e_2(x_a^{(1)})]^N\prod_{b=1}^{M_{2}}e_1(x_a^{(1)} -x_b^{(2)}) = 
\prod_{b=1}^{M_{1}}e_2(x_a^{(1)} -x_b^{(1)}), \nonumber\\
&& \prod_{b=1}^{M_{K-1}}e_{1}(x_a^{(K)} -x_b^{(K-1)}) = 
\prod_{b=1}^{M_K}e_2(x_a^{(K)} -x_b^{(K)}), \nonumber\\
&& E = \sum_{a=1}^{M_1} \frac{1}{2i}\ln e_2(x_a^{(1)}).\label{bareBA2}
\eea
With these equations, see also Ref.~\cite{tsveliknjp}, one can obtain thermodynamic observables in an exact manner for arbitrary even $M$. 

\subsection{Odd $M$. Detailed description of  $M=5$}

For the SO($M=2K+1$) group, the BA equations (up to the driving terms)  
can be extracted, for instance,  from Ref.~\cite{PCF}. 
The positions of the bulk and the impurity driving terms are determined by the same logic as before, that is by representation theory 
considerations and the $M \rightarrow M-2$ flow. 
For the SO$_2$($2K+1$) model, we suggest  the following bare BA equations, see also Ref.~\cite{tsveliknjp},
\bea
&& e_{1/2}(x_a^{(K)} -1/\bar\lambda)\prod_{b=1}^{M_{K-1}}e_1(x_a^{(K)} - x_b^{(K-1)}) =  \prod_{b=1}^{M_K}e_1(x_a^{(K)} - x_b^{(K)}),\nonumber\\
\nonumber &&\prod_{b=1}^{M_{p-1}}e_1(x_a^{(p)} - 
x_b^{(p-1)})\prod_{b=1}^{M_{p+1}}e_1(x_a^{(p)} - 
x_b^{(p+1)})\\ 
&&= \prod_{b=1}^{M_p}e_2(x_a^{(p)} - x_b^{(p)}),\quad p=2,\ldots,K-1,\nonumber\\ 
&& [e_2(x_a^{(1)})]^{N}\prod_{b=1}^{M_2}e_1(x_a^{(1)} - 
x_b^{(2)}) = \prod_{b=1}^{M_1}e_2(x_a^{(1)} - x_b^{(1)}),\nonumber\\ 
&& E = \frac{1}{2i}\sum_a \ln[e_k(x_a^{(1)})]. \label{ON}
\eea
The impurity is in the spinor representation, and its driving term is 
in the first equation. For $M=5$, we shall see that this  is consistent with the 
flow $M=5 \to M=3$ driven by a single-component Zeeman field.  

To illustrate the case of odd $M$, we now analyze Eqs.~(\ref{ON}) for $K=2$, i.e., for
the group SO(5).  The corresponding equations for the  densities are 
\bea
&& s*A_{n,2} = \label{first} \tilde\s_n + A_{nm}*\s_m - [s(\omega/2)*A_{2n,m}(\omega/2)]*\rho_m,\\
&& \frac{1}{N}[a_n(\omega/2)]e^{i\omega/\bar\lambda} =\label{last}\\
&&  -[s(\omega/2)*A_{n,2m}(\omega/2)]*\s_m + 
\tilde\rho_n + [A_{nm}(\omega/2)]*\rho_m. \nonumber
\eea
From Eq.~(\ref{last}), we can then derive the TBA equations.
There are two types of energies, where $\phi_n$ $(\xi_n$) is related to $\rho_n$ ($\s_n$). With $n=0,1,2,\ldots$, we obtain
\bea
&& \phi_{2n+1} = s_{1/2}*\ln\left[(1+e^{\phi_{2n}})(1+e^{\phi_{2n+2}})\right],\nonumber\\
&& \phi_{2n} = - \delta_{n,2}e^{-2\pi x/3} + s_{1/2}*
\ln\left[(1+e^{\phi_{2n-1}})(1+e^{\phi_{2n+1}})\right] - \nonumber\\
&& \frac{s_{1/2}}{1-s}*\Big[s_{1/2}C_{2n,m}*\ln(1+e^{\phi_m}) + 
C_{2n,m}*\ln(1+e^{\xi_m})\Big],\nonumber\\
&&-\ln(1+e^{-\xi_n}) = -\frac{C_{nm}}{1-s}*\ln(1+e^{\xi_{m}}) -\nonumber\\
&& \frac{s_{1/2}*C_{nm}}{1-s}*\ln(1+ e^{\phi_{2m}}) - 
\delta_{n,2}e^{-2\pi x/3}.\label{F5}
\eea
The impurity free energy reads
\bea
  F_{\rm imp} &=& - T\int d x \Big\{s_{3/2}[x + \frac{3}{2\pi}\ln(T_K/T)]\ln(
1 + e^{\phi_2(x)}) \nonumber\\
&+& s_{1/2}[x + \frac{3}{2\pi}\ln(T_K/T)]\ln(1 + e^{\phi_1(x)}) \nonumber\\
&+&  s_{1/2}*s_{3/2}[x + \frac{3}{2\pi}\ln(T_K/T)]\ln(1+ e^{\xi_1(x)})\Big\},\label{F5a}
\eea 
where $s_{n/2} = s(n\omega/2)$.  
 
We now address the flow $M=5 \rightarrow M=3$ with just one non-zero
Zeeman field component $h_{12}=h_0$, putting $T=0$ for simplicity. 
The nonvanishing densities are $\s_{2}$, $\tilde\s_2$, $\rho_{4}$, and
$\tilde\rho_4$, with the corresponding equations 
\bea
&& \frac{1}{2\cosh(\omega/2)} = [A_{2,2}]^{-1}*\tilde\s_2 +\s_2 - \left[
\frac{\cosh(\omega/4)}{\cosh(\omega/2)}\right]*\rho_{4},\label{gs1}\\
&& \frac{e^{i\omega/\bar \lambda}}{N}\frac{\tanh(\omega/4)}
{\sinh(\omega)}= \left[A_{4,4}(\omega/2)\right]^{-1}*
\tilde\rho_{4} + \nonumber \\ && +  \rho_{4}  -  \left[\frac{1}{2\cosh(\omega/4)}\right]*\s_{2}. \label{gs}
\eea
In the absence of the Zeeman field, $\s_2,\rho_{4} \neq 0$ 
on the entire real axis, and $\tilde\rho_{4} = \tilde\s_2 =0$.  
In the opposite case of large $h_0$, however,  $\tilde\s_{2}$ is significant and 
corresponds to the progressive emptiness of $\s_2$. Now $\s_2(x) \neq 0$ 
only at $x<B$, where $B$ is determined by the Zeeman field. 
As a result, the asymptotics of $\rho_{4}(x)$ at $+\infty$,
which implies the low-energy behavior of the free energy, 
is determined by Eq.~(\ref{last}).  Here we can approximate 
\be
\left[\frac{1}{2\cosh(\omega/2)}\right]*\s_{2} \approx A e^{-2\pi x} ,
\quad 
A =\int_{B}^{-\infty}d y e^{2\pi y}\s_2(y).
\ee 
In the end, we find that Eq.~(\ref{gs}) coincides with the equation for the 
SU$_{4}$(2) Kondo model with impurity spin $S=1/2$, corresponding
to the SO(3) topological Kondo effect.  
This once more illustrates the flow $M \to M-2$ induced by
lowering temperature below $T_h$.

Equation (\ref{ON}) also allows one to calculate  the ground-state impurity entropy.  Solving Eqs.~(\ref{F5}) for $x\to -\infty$, we find that
 $\xi_2, \phi_4 \to -\infty$.  As a consequence, the equations for 
$\phi_{1,2,3}$ and $\xi_1$ decouple from the rest. Their solution is given by
\be
e^{\phi_1} =e^{\phi_3} = 3/2,\quad 1+e^{\phi_2} = e^{2\phi_1},\quad e^{\xi_1} = 1/3.
\ee
Substituting this into Eq.~(\ref{F5a}), we find $S_{\rm imp} = \ln\sqrt 5$, in accordance with the result quoted in Sec.~\ref{sec2}.

\section{Conclusions}

To conclude, we have formulated a Bethe ansatz solution for the SO$(M)$ topological Kondo problem realized by a mesoscopic superconducting island coupled to external leads via $M>2$ Majorana fermions. In our previous paper \cite{altland3}, we reported that in this model,  the Majorana spin non-locally encoded by the Majorana fermions exhibits rich and observable dynamics characterized by nonvanishing multi-point correlations and nonperturbative crossovers between different non-Fermi liquid Kondo fixed points.  The Bethe ansatz results provided in the present work support these conclusions and provide a nonperturbative approach to the model spectrum and its thermodynamics. 

\ack

We thank A.A. Nersesyan and V. Kravtsov 
for valuable discussions, and acknowledge financial support by 
the SFB TR12 and the SPP 1666 of the DFG, a Royal Society URF, and the
DOE under Contract No.~DE-AC02-98CH10886.

%\appendix
%\section{}

\end{document}